# Doppler Shifted Unruh Radiation: The Non-Relativistic Spectrum

Jeffrey S. Lee[1,3]
Gerald B. Cleaver[1,2]
Felix Yu[3]

[1]Early Universe Cosmology and Strings Group, Center for Astrophysics, Space Physics & Engineering Research, Baylor University, One Bear Place, Waco, TX 76798

[2]Department of Physics, Baylor University, One Bear Place, Waco, TX 76798

[3]Crescent School, 2365 Bayview Avenue, Toronto, ON M2L 1A2

Jeff_Lee@Baylor.edu
FelixYu@crescentschool.org
Gerald_Cleaver@Baylor.edu

## 1. Abstract

In this note, the non-relativistic Doppler spectrum of the Unruh radiation from an accelerating mass is determined. Additionally, the scattering of thermal bath photons off an accelerating electron in an electromagnetic field, as seen by a non-laboratory frame inertial observer, is examined.

## 2. Introduction

The Unruh radiation field emitted by an accelerating charge has been well established in the literature. This paper considers a non-relativistic inertial observer moving in an arbitrary direction to an arbitrary photon in the radiation field, and invokes Doppler shifting and temperature inflation to account for the effects as seen by an observer who is not necessarily in the laboratory frame. The Doppler shifted, frequency-dependent Unruh energy density in the observer frame is calculated.

Additionally, the temperature-inflated and Doppler shifted contribution of photon field scattering off the accelerating electron is determined. It is further shown that if the charge is accelerated by an electric field in excess of a simple multiple of the quantum electrodynamic critical field strength, the Unruh radiation power reduces to the Doppler shifted Larmor radiation power.

## 3. The Doppler Shifted Unruh Radiation Spectrum

The thermal radiation bath temperature in the laboratory frame for an accelerating particle, as determined by Unruh, was shown to be [1]:

$$T_o = \frac{\hbar a}{2\pi c k} \qquad (1)$$

where $\hbar$ is the reduced Planck's constant, and $k$ is the Boltzmann constant. Eq. (1) has a demonstrated practical value in accelerator physics because quantum fluctuations of polarized synchrotron radiation allow storage ring electrons to only asymptotically approach complete polarization, even though they emit completely polarized synchrotron radiation [2], [3].

In the laboratory frame of an accelerating charge, the Unruh radiation spectrum is taken to be an isotropic photonic thermal bath, and the corresponding frequency-dependent energy density is given by the Planck blackbody expression,

$$\frac{dU_{\text{Unruh}}}{d\nu_o} = \left(\frac{8\pi}{c^3}\right) \frac{h\nu_o^3}{\exp\left(\frac{h\nu_o}{kT_o}\right) - 1} \quad (2)$$

where $\nu_o$ is the radiation proper frequency, $T_o$ is the proper temperature of the thermal bath, and $h$ is Planck's constant.

For an observer moving at non-relativistic speed $V=v/c$ in an arbitrary direction $\theta$ with respect to the direction of an individual photon, the non-relativistic Doppler shift is

$$\frac{\Delta\nu}{\nu_o} = -V\cos\theta \quad (3)$$

Therefore,

$$\nu = \nu_o(1 - V\cos\theta) \quad (4)$$

Clearly, the direction of the observer will be from $0°$ to $180°$ with respect to individual photons within the thermal bath. Additionally, observers in frames other than the rest frame of a heat source will observe a velocity-dependent inverse temperature, as discussed by Lee & Cleaver [4]. Although numerous unresolved issues surrounding relativistic thermodynamics are abound in the literature [4], [5], [6], [7], [8], [9], [10], [11], the *effective temperature* $T_{\text{eff}}$, in terms of the rest frame temperature $T_o$, is given by:

$$T_{\text{eff}} = \frac{T_o\sqrt{1-V^2}}{1-V\cos\theta} \quad (5)$$

The effects of temperature inflation are generally small at non-relativistic velocities; for example, at $V = 0.01$, an observer will experience a maximum effective temperature $(\theta = 0)$ that is only 1% greater than the rest frame temperature. However, temperature inflation is included here for completeness and to account for the predicted larger energy densities at lower frequencies. In the laboratory frame and without accounting for temperature inflation,

$$\frac{dU_{Unruh}}{d\nu} = \left(\frac{8\pi}{c^3}\right)\frac{h\nu^3}{\exp\left(\frac{h\nu}{kT}\right)-1} \tag{6}$$

Combining eqs. (1), (4), (5), and (6) results in the non-relativistic Doppler shifted, frequency-dependent Unruh energy density in the observer frame, which is given in eq. (7) and plotted in Figure 1.

$$\frac{dU_{Unruh}}{d\nu} = \left(\frac{8\pi}{c^3}\right)\frac{h\nu_o^3(1-V\cos\theta)^3}{\exp\left[\frac{4\pi^2 c\nu_o}{a}\frac{(1-V\cos\theta)^2}{\sqrt{1-V^2}}\right]-1} \tag{7}$$

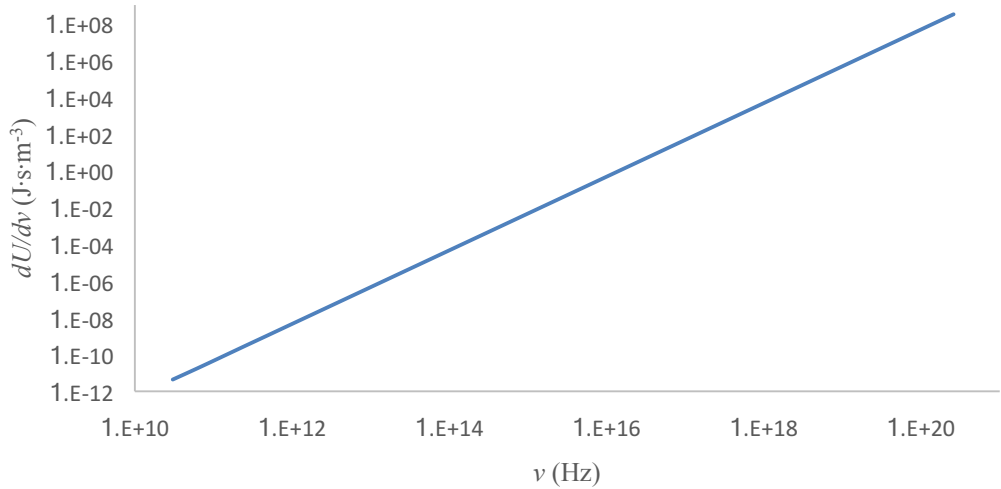

Figure 1: The non-relativistic Doppler shifted, frequency-dependent Unruh energy density in the observer frame versus frequency. The angle between the observer's velocity vector and emitted photons of interest is 30°. The observer's velocity is $0.1c$, and the acceleration of the Unruh radiation-emitting electron is $10^{35}$ m/s².

For frequencies of $10^{10}$-$10^{20}$ Hz (wavelengths from 1 cm to 1 pm), temperature inflation causes a mean increase in the Unruh energy density of 19.3% above the Unruh energy density for which temperature inflation is not considered. Variations in this increase are between 19% and 23% for the conditions considered in Figure 1, are approximately constant throughout most of the above-mentioned frequency range, and increase sharply at lower frequencies (see Figure 2).

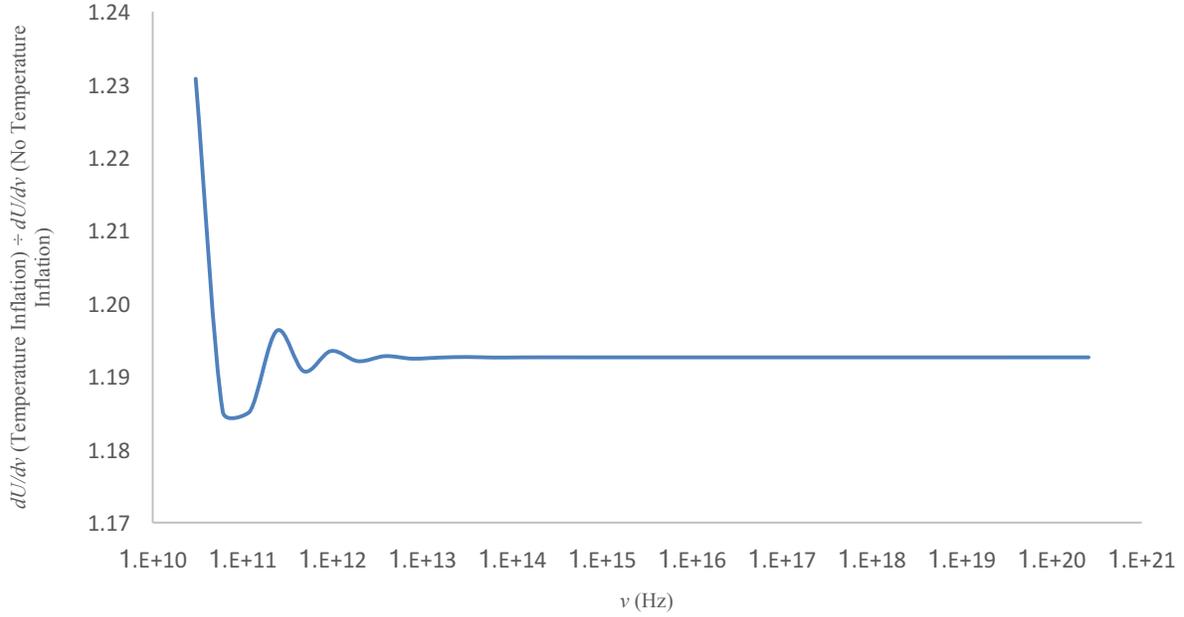

Figure 2: Ratio of non-relativistic Doppler shifted, frequency-dependent Unruh energy densities, with and without temperature inflation, versus frequency. The angle between the observer's velocity vector and emitted photons of interest is 30°. The observer's velocity is $0.1c$, and the acceleration of the Unruh radiation-emitting electron is $10^{35}$ m/s².

### 4. The Flux Scattered off the Electron

The scattering of photons off the accelerating electron results in an additional contribution to the radiation flux Φ, as seen by an observer in the laboratory frame [1].

Thus,
$$\Phi = cU \tag{8}$$

and,
$$\frac{dU_{Unruh}}{d\nu dt} = \frac{dU_{Unruh}}{d\nu} \sigma_{Thomson} \tag{9}$$

where $\sigma_{Thomson}$ is the familiar Thomson scattering cross-section given by $\sigma_{Thomson} = \frac{8\pi r_o^2}{3}$, in which $r_o = \frac{e^2}{mc^2}$ is the classical electron radius. Therefore, combining eqs. (7), (9), and the expressions for the Thomson scattering cross-section $\left(\sigma_{Thomson} = \frac{8\pi r_o^2}{3}\right)$ and the classical electron radius $\left(r_o = \frac{e^2}{mc^2}\right)$ results in the Doppler shifted isotropic radiation flux experienced by the electron and seen by a non-relativistic inertial observer.

$$\frac{dU_{Unruh}}{d\nu dt} = \left(\frac{64\pi^2 e^4}{3m^2 c^6}\right) \frac{h\nu_o^3 (1-V\cos\theta)^3}{\exp\left[\frac{h\nu_o}{kT_o}(1-V\cos\theta)\right]-1} \tag{10}$$

Integrating over the proper frequency yields the Unruh radiation power density,

$$\frac{dU_{Unruh}}{dt} = \left(\frac{64\pi^2 e^4}{3m^2 c^6}\right) \int_0^\infty \frac{h\nu_o^3 (1-V\cos\theta)^3}{\exp\left[\frac{h\nu_o}{kT_o}(1-V\cos\theta)\right]-1} d\nu_o \tag{11}$$

Evaluating eq. (11) yields,

$$\frac{dU_{Unruh}}{dt} = \left(\frac{8\pi^3 e^4 \hbar}{45 m^2 c^2}\right)\left(\frac{kT_o^4}{\hbar}\right)(1-V\cos\theta)^{-1} \tag{12}$$

where $\hbar$ is the Reduced Planck's constant. Written in terms of the acceleration in the observer's instantaneous rest frame and the classical electron radius, eq. (12) becomes:

$$\frac{dU_{Unruh}}{dt} = \left(\frac{\hbar r_o^2 a^4}{90\pi c^6}\right)(1-V\cos\theta)^{-1} \tag{13}$$

If the observer is at rest in the laboratory frame, then eq. (13) reduces to the familiar Hawking-Unruh relation [1].

$$\frac{dU_{Unruh}}{dt} = \frac{\hbar r_o^2 a^4}{90\pi c^6} \tag{14}$$

In the event that the acceleration-causing electromagnetic field is $E^*$ (given by eq. (15)), then the Unruh radiation power is given by eq. (16).

$$E^* = \sqrt{\frac{60\pi}{\alpha}} E_c, \tag{15}$$

$$\frac{dU_{Unruh}}{dt} = \left(\frac{2e^2 a^2}{3c^3}\right)(1-V\cos\theta)^{-1} \tag{16}$$

where $E_c = \frac{m^2 c^3}{e\hbar}$ is the quantum electrodynamic critical field strength (i.e. Schwinger Limit), and $\alpha$ is the fine structure constant. Eq. (16) is, expectedly, the classical Larmor radiation power being Doppler shifted for an inertial observer, which trivially reduces to eq. (17) for an observer in the laboratory frame.

$$\frac{dU_{\text{Unruh}}}{dt} = \left(\frac{2e^2 a^2}{3c^3}\right) \tag{17}$$

## 4. Summary

The non-relativistic Doppler spectrum of the Unruh radiation of an accelerating mass has been determined. Temperature inflation was accounted for by invoking the effective temperature. Expectedly, the non-relativistic Doppler shifted, frequency-dependent Unruh energy density in the observer frame was considered to be Planckian in configuration. A comparison was made of the non-relativistic Doppler shifted, frequency-dependent Unruh energy densities, with and without temperature inflation. This revealed that the effects of considering temperature inflation became significant only when considering lower frequency Unruh photons.

The additional radiation flux contribution, as seen by an inertial observer, resulting from the scattering of Unruh photons off an accelerating electron, was determined. From this, the Unruh power density was calculated. When the acceleration-causing electromagnetic field strength was a minimum multiple of the quantum electrodynamic critical field strength, the familiar expression for the Larmor radiation power emerged.